# Harmonic-seeded remote laser emissions in $N_2$-Ar, $N_2$-Xe and $N_2$-Ne mixtures: a comparative study

Jielei Ni [1,2], Wei Chu [1,2], Haisu Zhang [1,2], Chenrui Jing [1,2], Jinping Yao [1], Huailiang Xu [3,6], Bin Zeng [1,2], Guihua Li [1,2], Chaojin Zhang [1,4], See Leang Chin [5], Ya Cheng [1,7], Zhizhan Xu [1,8]

[1] *State Key Laboratory of High Field Laser Physics, Shanghai Institute of Optics and Fine Mechanics, Chinese Academy of Sciences, P.O. Box 800-211, Shanghai 201800, China*

[2] *Graduate School of the Chinese Academics of Sciences, Beijing 100039, China*

[3] *State Key Laboratory on Integrated Optoelectronics, College of Electronic Science and Engineering, Jilin University, Changchun 130012, China*

[4] *College of Physics and Electronic Engineering, Jiangsu Normal University, Xuzhou 221116, China*

[5] *Center for Optics, Photonics and Laser, Université Laval, Quebec City, Quebec G1V 0A6, Canada*

[6] *huailiang@jlu.edu.cn*

[7] *ya.cheng@siom.ac.cn*

[8] *zzxu@mail.shcnc.ac.cna*

**Abstract:** We report on the investigation on harmonic-seeded remote laser emissions at 391 nm wavelength from strong-field ionized nitrogen molecules in three different gas mixtures, i.e., $N_2$-Ar, $N_2$-Xe and $N_2$-Ne. We observed a decrease in the remote laser intensity in the $N_2$-Xe mixture because of the decreased clamped intensity in the filament; whereas in the $N_2$-Ne mixture, the remote laser intensity slightly increases because of the increased clamped intensity within the filament. Remarkably, although the clamped intensity in the filament remains nearly unchanged in the $N_2$-Ar mixture because of the similar ionization potentials of $N_2$ and Ar, a significant enhancement of the lasing emission is realized in the $N_2$-Ar mixture. The enhancement is attributed to the stronger third harmonic seed, and longer gain medium due to the extended filament.

1. Introduction

The investigation of lasing from nitrogen molecules can date back to the early 1960s [1,2]. In recent years, technological advances in ultrafast laser source have allowed to access filaments over a distance up to several kilometers [3-5], inspiring strong interest in the development of filament-assisted remote laser and its application in standoff spectroscopy [6-15]. Recent investigations have concentrated on molecular nitrogen laser initiated by amplified spontaneous emission (ASE) with the operation wavelength at ~337 nm or ~357 nm [6-12] and atomic oxygen laser by ASE at 845 nm [13]. The former is based on the electron recombination of ionized molecular nitrogen [6-9] or thermal electron collisional excitation [10-12], and the latter is realized by simultaneous dissociation of molecular oxygen and excitation of atomic oxygen using a picosecond 226 nm-laser beam [13].

Different from the ASE schemes mentioned above, we recently demonstrated a novel approach to realize switchable multi-wavelength laser in air [14] and simultaneous multi-wavelength-generated ultraviolet laser in carbon dioxide [15]. In our scheme, molecular nitrogen and carbon dioxide ions induced by ultrafast laser filamentation of femtosecond infrared laser pulses serve as the gain medium with the population inversion established in an ultrafast time scale comparable to the pump laser pulse. Besides, the lasing emissions originate from stimulated amplification seeded by self-generated harmonics of the pump laser during filamentation, which enables production of stronger laser emissions than those reported in ref. [6-12] and simpler operation than that in ref. [13]. In a broad sense, the discoveries of the strong harmonic-seeded remote laser emissions from $N_2$ and $CO_2$ gas media are clear evidences of the advantages of investigating femtosecond laser filamentation with intense, wavelength-tunable mid-infrared ultrafast sources. The research field is still in its infancy, but some exciting results have already been achieved [16-19].

In this paper, we systematically study the behaviors of the harmonic-seeded nitrogen laser inside a femtosecond laser-induced filament in nitrogen mixed with xenon, argon or neon. In different gases the clamped laser intensity inside the filament core would vary, depending on the nonlinear Kerr refractive index and the ionization potential. The highest clamped intensity is produced in the $N_2$-Ne mixture, which is followed by the $N_2$-Ar mixture and then the $N_2$-Xe mixture. It is reasonable to expect a higher yield of 391 nm lasing emission in the $N_2$-Ne mixture because of the higher clamped intensity. Interestingly, we observed that the enhancement in the $N_2$-Ar mixture is more remarkable. The mechanisms responsible for the observations will be discussed in details. Moreover, we also show in this study simultaneous laser emissions at 357 and 391 nm, or at 420 nm, 424 nm, and 428 nm by pump lasers with different wavelengths.

2. Experiment

The experimental setup is shown in Fig. 1. Wavelength-tunable mid-IR femtosecond laser pulses are generated by an optical parametric amplifier (OPA) (HE-TOPAS, Light Conversion, Inc.) pumped by a commercial Ti: sapphire laser system (Legend Elite-Duo, Coherent, Inc.) with a repetition rate of 1 kHz and a central wavelength of ~800 nm. The mid-IR pulse is focused into a vacuum chamber filled with different gas mixtures ($N_2$-Ne, $N_2$-Ar, or $N_2$-Xe) by a fused silica lens with a focal length of 24.3 cm. The focal length is measured at 1184 nm wavelength taking into account the 3-mm-thick window of the vacuum chamber. The diameter of the laser beam impinged on the lens is ~8 mm at $1/e^2$ of the energy fluence. Output signals are collimated by a lens with a focal length of 25 cm, and then sent into an imaging grating spectrometer (Shamrock 303i, Andor) with a 1200 grooves/mm grating. The entrance slit of the spectrometer is set to be 50 μm. Side fluorescence is imaged by a lens with a focal length of 4cm and collected by the same spectrometer (Shamrock 303i, Andor).

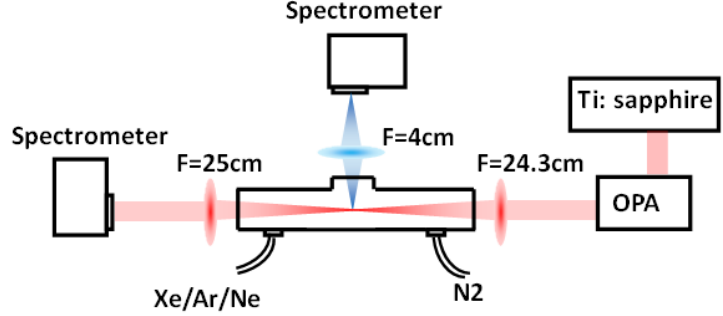

Fig.1. Schematic of the experimental setup.

In this experiment, nitrogen is mixed respectively with three different gases: A. xenon; B. neon; C. argon. The ionization potentials ($I_p$) and the second-order Kerr refractive index coefficients ($n_2$) of all the gases used are listed in Table 1 [20-23]. In these three different gas mixtures, the pump laser wavelength is fixed at 1184 nm with an average power of 960 mW. Fluorescence signals of ionized nitrogen molecules at 391 nm are recorded from the side of filament with 1 s exposure time, corresponding to 1000 laser shots. Forward 391nm-laser is attenuated to 1% and is recorded with 0.1s exposure time, corresponding to 100 laser shots.

Table 1. Ionization potentials and second-order Kerr refractive index coefficients (800 nm at 1 atm) of Xe, $N_2$, Ar, and Ne.

| Material | Xe | $N_2$ | Ar | Ne |
|---|---|---|---|---|
| $I_p$ (eV) | 12.13 | 15.58 | 15.76 | 21.56 |
| $n_2$ ($10^{-20}$cm$^2$/W) | 93[20,21] | 11[22,23] | 10[22,23] | 0.74[20,21] |

3. Results

In Fig. 2, we show the spectra of the 3$^{rd}$ harmonic as well as the amplified 391 nm signal in the forward direction of the pump laser in the $N_2$-Xe (2a), $N_2$-Ar (2b), and $N_2$-Ne (2c) mixtures, respectively. The 391 nm signals presented in Figs. 2(d), 2(e) and 2(f) are obtained by subtracting the 3$^{rd}$ harmonic background in these three mediums. In these cases, the 3$^{rd}$ harmonic background was obtained firstly by removing the lasing spectral values in a narrow range from 391 nm to 393 nm from the broad 3$^{rd}$ harmonic spectrum and then filling it up using the interpolation method. As can be seen from Fig. 2(d), the injection of xenon immediately quenches the 391 nm lasing emission. In contrast, it can be clearly observed from Fig. 2(f) that the 391 nm signal increases slowly as the neon pressure increases. Fig. 2(e) demonstrates the result of the $N_2$-Ar mixture, from which distinct enhancement of the 391 nm signal can be observed at 200 mbar argon pressure and saturation above 200 mbar. To compare the enhancement effect of 391 nm signals in the $N_2$-Ar mixture with that in the $N_2$-Ne mixture, we plot the intensities of the 391 nm laser signals obtained in Figs. 2(e-f) as a function of the pressure of the inert gas added into the N₂ gas, as shown in Fig. 2(g). Clearly, the enhancement effect is much more remarkable in the N₂-Ar mixture than in the N₂-Ne mixture.

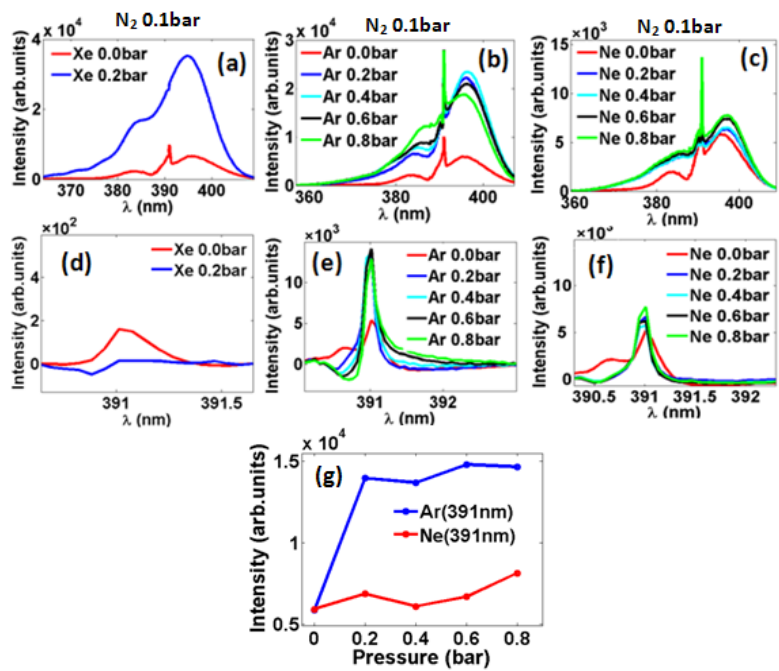

Fig. 2 Measured spectra of the 3rd harmonic (a, b, c) and pure 391 nm lasing signals (d, e, f) in the forward direction in N2-Xe mixture (a, d), an N2-Ar mixture (b, e) and N2-Ne mixture (c, f) using pump wavelength at 1184 nm. The pressure of nitrogen is fixed at 0.1bar. Comparison of the 391 nm peak intensity in N2-Ar gas mixture and N2-Ne gas mixture is shown in (g).

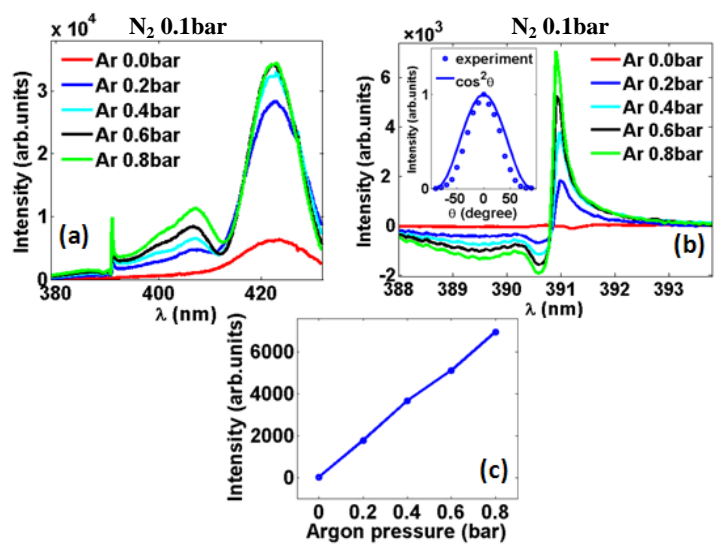

Fig. 3 Measured spectra of the 3$^{rd}$ harmonic (a) and pure 391 nm lasing signals (b) in the forward direction in N2-Ar mixture using pump wavelength at 1258 nm. The pressure of nitrogen is fixed at 0.1bar. The 391 nm peak intensity as a function of argon pressure is depicted in (c). Inset in (b): Polarization property of the 391 nm laser.

We notice that the enhancement introduced by argon can also exist when we tune the pump wavelength to 1258 nm at which the lasing signal at 391 nm is very weak (barely observable) in pure nitrogen. This is illustrated in Figs. 3(a-b). To verify that the 391 nm emission is generated by amplification of the harmonic seed rather than by spontaneous emission or amplified spontaneous emission (ASE), we measured its polarization property as presented in the inset of Fig. 3(b). A nearly perfect linear polarization is obtained as shown in the inset of Fig. 3(b), indicating that the 391 nm signal results from amplified seed action because both spontaneous emission and ASE show isotropic polarization. To demonstrate the enhancement of 391 nm laser emission in the $N_2$-Ar mixture, the 391 nm intensity is depicted in Fig. 3(c) as a function of the argon pressure, which gives an almost linear dependence without saturation.

## 4. Discussion

In order to interpret the different behaviors of the 391 nm laser emission, we describe the output of the 391 nm lasing signal from the gas mixture by the well-known small signal gain equation for simplicity, $I_{laser} = I_{seed} e^{gL}$ [24], where $L$ stands for the length of a gain medium (the filament length in our experiment), $I_{seed}$ is the seed intensity (the 3$^{rd}$ harmonic intensity in our experiment), and $g$ is the gain coefficient of the medium. The filament length in different gas mixtures for a collimated beam can be roughly evaluated by Marburger's law [25] which is used for the calculation of the onset of the filament, $L_c = \dfrac{0.367 k a_0}{\sqrt{[(P_{in}/P_{cr})^{1/2} - 0.852]^2 - 0.0219}}$, with $k$ and $a_0$ being the wave number and the radius of the beam profile at the 1/e level of intensity, and $P_{in}$ and $P_{cr}$ being the input power of the laser pulse and the critical power, respectively. The critical power $P_{cr}$ can be expressed as $P_{cr} = \dfrac{3.72 \lambda_0^2}{8 \pi n_0 n_2(P)}$ [25] with $n_2(P)$ being proportional to pressure $P$ [20] . In the case of a convergent beam with a lens ($f$) the onset of the filament is modified to be $\dfrac{1}{L_{c,f}} = \dfrac{1}{L_c} + \dfrac{1}{f}$ [25], and thus the gain medium length $L$ is expressed as $L = f - L_{c,f} = \dfrac{f^2}{f + L_c}$ [6]. The gain coefficient $g = \Delta n \times \delta_{21}$ depends on the population inversion $\Delta n$ and the stimulated emission cross section $\delta_{21}$ [24]. In our experiment, the population inversion is launched by a cooperative process of strong field ionization and excitation from the ground state ($X^2\Sigma_g^+$) of $N_2^+$ to the excitation state ($B^2\Sigma_u^+$) by the fundamental infrared light and its 3$^{rd}$ and 5$^{th}$ harmonics [14,15]. It is therefore strongly dependent on the ionization rate of $N_2$ and accordingly the clamped intensity in the filament.

In different kinds of gases, the clamped intensity $I$ varies following the relationship $n_2 I = N_e(I)/(2 N_{crit})$ [26]. Here, $N_e(I)$ is the electron density induced by multiphoton/tunnel ionization and thus strongly depends on the ionization potential $I_p$ of the atoms/molecule in the filament, and $N_{crit}$ is the critical plasma density given by $N_{crit} = \varepsilon_0 m \omega^2 / e^2$ ($m$, $e$, $\omega$ are the electron mass, elementary charge, and the central frequency of the laser field, respectively). Since both $n_2$ and $N_e$ are proportional to pressure, the clamped intensity is independent on pressure. According to this formula, the clamped intensity in nitrogen is higher than that in xenon, lower than that in neon, and comparable with that in argon. In the case of a gas mixture, the clamped intensity depends on the concentration ratio of different gas species in the mixture.

In the $N_2$-Xe mixture, the clamped intensity will decrease with the increase of xenon pressure and gradually approach the value obtained in pure xenon at high xenon pressure. To confirm this, we measure fluorescence intensity of 391 nm from the side of the filament in Fig. 4(a). The 391 nm fluorescence stems from the radiative decays of the excited state $B^2\Sigma_u^+$ of $N_2^+$ [27] and thus depends on the ionization rate of nitrogen and correspondingly the clamped intensity inside the filament. Therefore, the clamped intensity can be indicated from the fluorescence intensity at 391 nm. In Fig.

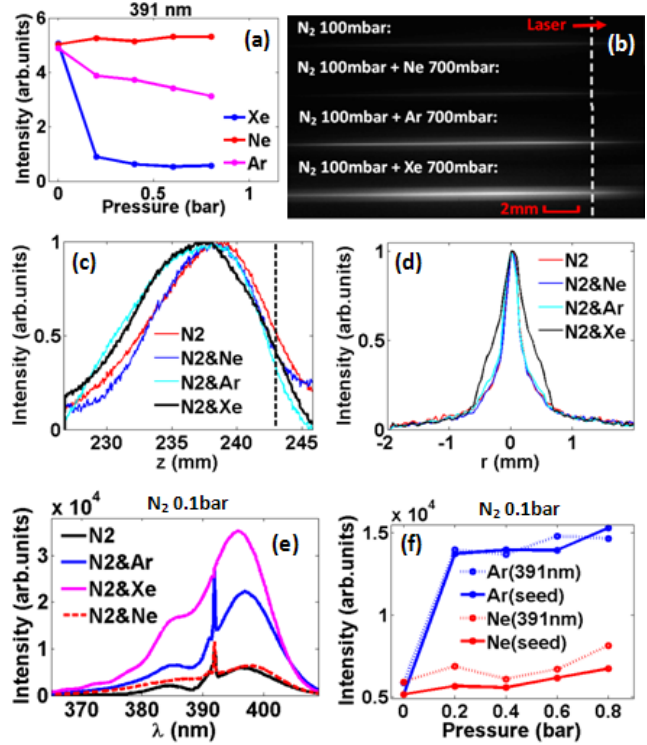

Fig. 4 (a) Experimentally measured fluorescence intensities at 391 nm in the three gas mixtures. The pressure of nitrogen is fixed at 0.1bar. (b) Images of femtosecond filaments in the three gas mixtures and pure nitrogen. (c) Normalized plasma luminescence intensity distribution inside the filaments in (b) along the direction of laser propagation. (d) Normalized plasma luminescence intensity inside the filaments in (b) in lateral direction. (e) Forward 3rd harmonic spectra in pure nitrogen (100mbar), $N_2$-Xe gas mixture (100:200mbar), $N_2$-Ar gas mixture (100:200mbar) and $N_2$-Ne gas mixture (100:200mbar). (f) Comparison of the 391 nm lasing intensity and its seed intensity in $N_2$-Ar gas mixture and $N_2$-Ne gas mixture.

4(a), the 391 nm fluorescence decays rapidly with the increase of xenon pressure and stabilizes at high xenon pressure, suggesting the clamped intensity inside the filament sharply decreased as the xenon pressure increased and finally reach the value in pure xenon. Note that the filament width also impacted on the fluorescence intensity. In Fig. 4(b) we show four typical images of the filaments recorded over ~100 shots in pure nitrogen (100 mbar) as well as different types of gas mixtures, which were captured by a CCD camera (WinCamd-UCD23) from the side of the filament. The response of the CCD covers from 350 nm to 1150 nm with comet tail elimination for λ > 900 nm. No filter was used. The arrow indicates the propagation direction of the laser pulses. The dashed line shows the position of the geometric focus. Normalized filament luminescence in both lateral and longitudinal directions is depicted in Figs. 4(c) (the geometrical focus is located at z = 243 mm as shown by the dash line) and (d), respectively. From Fig. 4(d) we can see that the filament in the N2-Xe mixture gets much broader than that in the pure nitrogen, which may reflect the ionization potential of different gases, that is, the lower ionization potential is, the broader the filament diameter becomes. We speculate that the central bright line in $N_2$-Xe mixture is from the filament core and the weak luminescence is from the energy reservoir. This broader filament diameter will result in a larger number of fluorescent nitrogen ions and in turn stronger fluorescence intensity. Judging from the decay curve in Fig. 4(a), the influence of the broadened filament width on the fluorescence signal should be relatively weak as compared with the influence of the decreasing clamped intensity in the filament.

The sharply reduced clamped intensity will lead to a great reduction in the generation of nitrogen ions, and strongly suppress the population inversion $\Delta n$. Although the high nonlinearity of xenon enables efficient generation of the 3$^{rd}$ harmonic which is clearly shown in Fig. 4(e), and a larger volume of filament that can be clearly observed in Figs. 4(b-d), the strongly suppressed population inversion leads to the quenching of the 391 nm lasing emission, as demonstrated in Figs. 2(a) and 2(d).

In the N$_2$-Ne mixture, the clamped intensity is expected to increase with the increase of neon pressure and gradually approach the value obtained in pure neon. However, within the pressure range of our experiment, the increase of clamped intensity is small, which is evidenced by the slight increase of the 391 nm fluorescence in Fig. 4(a). This is because the high ionization potential of neon makes it hard to be ionized, so that its contribution to the electron density is negligible. To exclude the influence of filament width on the fluorescence intensity, we note that the filament is mainly supported by nitrogen molecules in this case, so that the change in the filament size is not significant since the pressure of nitrogen is fixed. This is also confirmed by the lateral and longitudinal plasma luminescence profiles in Figs. 4(c-d).

The slightly increased clamped intensity (in N$_2$-Ne mixture) consequently leads to the slight increase of the 391 nm laser emission. The reasons are as follows. First of all, the increased clamped intensity will promote the ionization of nitrogen and the population inversion $\Delta n$. Secondly, note that the 3$^{rd}$ harmonic in a filament scales as $I_{3w} \propto \left(\chi^{(3)}_{Ne} + \chi^{(3)}_{N2}\right)^2 I_w^3$ (where $\chi^{(3)} = 4\varepsilon_0 c n_0^2 n_2(P)/3$) [25,28], therefore the increase of both the neon pressure and the clamped intensity can contribute to the enhancement of the 3$^{rd}$ harmonic intensity which serves as the seed.

In the N$_2$-Ar mixture, the clamped intensity should remain nearly the same with changing gas pressure of argon because of the similar second order Kerr refractive index coefficients ($n_2$) and ionization potentials ($I_p$) values of argon and nitrogen, according to the formula $n_2 I = N_e(I)/(2N_{crit})$ [26]. The slight decrease of the 391 nm signal shown in Fig. 4(a) comes from the collisional quenching due to the increasing pressure. As a result of the relatively stable intensity in the filament, the population inversion $\Delta n$ is expected to maintain nearly the same. In addition, since $n_2$ increases with pressure, the onset of the filament is shifted away from the focus of the lens at high pressure of argon according to Marburger's law, giving rise to a longer filament. This can also be observed in Figs. 4(b) and 4(c). More importantly, the injection of argon beyond 0.2 bar promotes the 3$^{rd}$ harmonic intensity as a strong and almost constant seed as can be seen from Fig. 2(b) and 2(e). Both the extended filament length and stronger 3$^{rd}$ harmonic seed can lead to the enhancement of the 391 nm lasing emission.

To further clarify the role of the 3$^{rd}$ harmonic intensity in the enhancement of the 391 nm laser emission, we compare the 391 nm laser peak intensity and its seed intensity in Fig. 4(f) in N$_2$-Ar mixture and N$_2$-Ne mixture. In this case, the 391 nm laser peak intensity ($I_{output} - I_{3\omega}$) in the two gas mixtures is the peak intensity in Fig. 2(e) and Fig. 2(f), respectively. The seed intensity is the interpolation value of the 3$^{rd}$ harmonic intensity at the peak wavelength of the 391 nm lasing emission. As we can see from Fig. 4(f), the 391 nm laser peak intensity closely follows the seed intensity. In the N$_2$-Ar mixture, it stops growing at pressures above 200 mbar due to saturation of the seed inside the core of the filament. We ascribe the saturation of the 3$^{rd}$ harmonic to the ring structure of the 3$^{rd}$ harmonic beam profile at high pressure [18,19,29]. In this case, the increase of the 3rd harmonic with argon pressure will be contained in the ring due to plasma defocusing, leading to the saturation of the 3rd harmonic inside the core. Since the 391 nm laser emission emerges from the core of the filament, the saturation of the 3rd harmonic together with the nearly constant intensity of the pump inside the filament core inevitably results in the saturation of the 391 nm laser emission.

Comparing the seed and 391 nm laser signals obtained in $N_2$-Ar with those obtained in $N_2$-Ne as illustrated in Fig. 4(f), the laser emission at 391 nm is much stronger in the $N_2$-Ar mixture than that in the $N_2$-Ne mixture. This is ascribed to the much stronger seed and the longer filament length as shown in Figs. 4(b) and 4(c).

## 5. Simultaneous multi-wavelengths harmonic seeded $N_2^+$ remote laser

In the early work in Ref [14], the 5th harmonic seeded $N_2^+$ remote laser is obtained separately at 330, 357, 391, 428, and 471 nm by tuning the pump laser wavelength. In the present scheme, multi-wavelengths can be simultaneously accomplished in pure nitrogen by tuning the pump laser wavelength properly. This can be clearly seen in Fig. 5(a) in which the two laser signals at 357 nm and 391 nm are both seeded by the 5th harmonic of the infrared light at 1753 nm wavelength. In Fig. 5(b) we also demonstrate the simultaneous line emissions at 420 nm, 424 nm, and 428 nm in pure nitrogen, seeded by the 3rd harmonic of the infrared light at 1260 nm. The three laser emissions are attributed to the ($v'=2 \to v=3$) transition (420 nm), the ($v'=1 \to v=2$) transition (424 nm) and the ($v'=0 \to v=1$) transition (428 nm) between $B^2\Sigma_u^+(v'=0,1,2)$ and $X^2\Sigma_g^+(v=1,2,3)$ states in ionized nitrogen molecules.

This is not reported in Ref. [14]. The simultaneous multi-wavelength laser emissions mentioned above stem from the enhanced spectral broadening of the harmonic seed in a long filament [25], because in such case, the spectrum of the harmonic seed enables covering the wavelengths of multiple lasing lines [15].

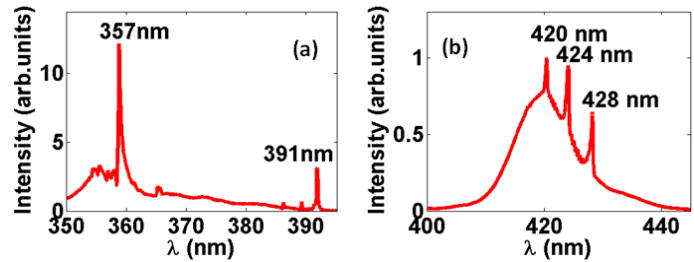

Fig. 5 (a) Multi-wavelengths remote laser at 357 nm and 391 nm in pure nitrogen. (b) Multi-wavelengths remote laser at 420 nm, 424 nm, and 428 nm in pure nitrogen.

## 6. Conclusion

In conclusion, we have systematically investigated the behavior of the 3rd harmonic seeded remote lasers in different types of gas mixtures. We show that although the signal intensity can be slightly increased in the $N_2$-Ne mixture due to the increased clamping intensity, the enhancement in the $N_2$-Ar mixture is much more significant. The enhancement mainly stems from the increase of both the 3rd harmonic intensity and the filament length with increased argon pressure, as the ionization rate of nitrogen almost remain unchanged owing to the intensity clamping. Simultaneous laser emissions have also been achieved in pure nitrogen.

The experiment carried out in this work not only demonstrates enhancement of the remote lasing signal using gas mixtures, but also provides a new approach for gaining deep insight into the underlying physics of the establishment of population inversion. In generally, inside a filament induced in a certain type of gas medium, it is not easy to tune the peak intensity of the pump laser in a wide range because of the well-known intensity clamping [25,26]. This creates difficulty in studying the intensity dependence of the remote lasing signal. However, with different combination of gas mixtures, we are able to significantly vary the peak intensity in a filament, taking advantage of the different ionization potentials of different inert atoms or molecules. Therefore, the use of gas mixture provides us an efficient way for tuning the pumping conditions in a filament that is not reachable in a pure nitrogen gas, which benefits the investigation on the mechanism behind population inversion. The results obtained above clearly shows that the key factors for efficiently achieving the harmonic-seeded remote lasing are high pump intensity and strong self-generated harmonic seed. Only in the nitrogen-argon mixture, both conditions are satisfied, thus a dramatic enhancement of the lasing signal at 391 nm wavelength has been observed.

**Acknowledgement**

This work is financially supported by National Basic Research Program of China (Grant 2011CB808102), National Natural Science Foundation of China (Grant Nos. 11134010, 11074098, 60825406, 10974213), NCET- 09-0429 and the fundamental research funds of Jilin University.